\documentclass[prl,twocolumn,showpacs,aps]{revtex4-1}

\usepackage{amsmath, amsfonts, amssymb, mathrsfs}
\usepackage{amssymb}
\usepackage{txfonts}
\usepackage{graphicx}
\usepackage{epsfig}
\usepackage{enumerate}
\usepackage{caption}
\usepackage{subcaption}

\usepackage{slashed}
\usepackage{ulem}

\usepackage{ifpdf}
\ifpdf
\usepackage{epstopdf}
\fi

\usepackage[unicode]{hyperref}
\hypersetup{
	pdftitle={Viscosity},
	pdfauthor={Nian, Pando Zayas},
	pdfsubject={Viscosity},
	colorlinks=true,
	linkcolor=blue,
	citecolor=blue,
	filecolor=black,
	urlcolor=blue
}

\def\url#1{}

\usepackage{bm}
\usepackage{color}

\newcommand{\be}{\begin{equation}}

\newcommand{\ee}{\end{equation}}

\newcommand{\bea}{\begin{eqnarray}}

\newcommand{\eea}{\end{eqnarray}}

\begin{document}
\hfill LCTP-20-29

\title{Retarded Green's Function from Rotating AdS Black Holes: Emergent CFT$_2$ and Viscosity}
\author{Jun Nian$^{1}$}
\email{nian@umich.edu}
\author{Leopoldo A. Pando Zayas$^{1, 2}$}
\email{lpandoz@umich.edu}
\affiliation{$^1$ Leinweber Center for Theoretical Physics, University of Michigan, Ann Arbor, MI 48109, USA}
\affiliation{$^2$ The Abdus Salam International Centre for Theoretical Physics, 34014 Trieste, Italy}

\begin{abstract}
Using the AdS/CFT correspondence we consider the retarded Green's function  in the background of rotating near-extremal AdS$_4$ black holes.  Following  the canonical AdS/CFT dictionary into the asymptotic boundary we get a CFT$_3$ result. We also take a new route and zoom in on the near-horizon region, blow up this region and show that it yields a CFT$_2$ result. We argue that the decoupling of the near-horizon region  is akin to the decoupling of the near-throat region of a D3-brane, which led to the original formulation of the AdS/CFT correspondence, thus implying that the Kerr/CFT correspondence follows as a decoupling of the standard AdS/CFT correspondence applied to rotating black holes.  As a byproduct, we compute the shear viscosity to entropy density ratio for the strongly coupled boundary CFT$_3$, and find that it violates the $1 / (4 \pi)$ bound.
\end{abstract}

\maketitle

{\bf Introduction.---}  One of the conceptual legacies of the AdS/CFT correspondence  \cite{Maldacena:1997re, Gubser:1998bc, Witten:1998qj} is its geometrization of the renormalization group flow in quantum field theories.  The interpretation of the radial direction of the asymptotically AdS spacetime as  the renormalization group scale   has been widely discussed \cite{Akhmedov:1998vf,deBoer:1999tgo, Bianchi:2001de, Bianchi:2001kw, Faulkner:2009wj, Heemskerk:2010hk, Faulkner:2010jy}. Most of the existing literature, however, has focused on the holographic counterpart of the field theory living at the asymptotic boundary and in spherically symmetric situations. In this manuscript we explore some implications of considering a rotating asymptotically AdS black hole background and an emergent IR CFT$_2$. 

The emergence of a Virasoro algebra as asymptotic symmetries in the near-horizon region of extremal Kerr black holes in asymptotically flat spacetimes is at the heart of the Kerr/CFT correspondence \cite{Guica:2008mu};  through the Cardy formula, this algebra microscopically reproduces the Bekenstein-Hawking entropy.  This correspondence, originally formulated in the near-horizon region of an extremal asymptotically flat Kerr black hole, has been generalized to near-extremal asymptotically flat and asymptotically AdS Kerr-Newman black holes \cite{Lu:2008jk, Chow:2008dp, Bredberg:2009pv, Hartman:2009nz, Chen:2010bh}. The question we pursue in this manuscript is: {\it what are the implications of embedding the Kerr/CFT correspondence in a larger AdS/CFT context?} We will see that this embedding provides a framework for a UV completion and justifies the Kerr/CFT correspondence as a stand-alone conjecture.

An important motivation to revisit the status of the Kerr/CFT correspondence in the larger context of AdS/CFT embeddings comes from recent developments in the physics of rotating, electrically charged, asymptotically AdS black holes. These black holes in dimensions four, five, six and seven,  have recently been provided a microscopic foundation for their Bekenstein-Hawking entropy  using the superconformal index of the dual CFT's \cite{Cabo-Bizet:2018ehj, Choi:2018hmj, Benini:2018ywd, Choi:2019miv, Kantor:2019lfo, Nahmgoong:2019hko, Choi:2019zpz, Nian:2019pxj}. Alternatively, a universal and unifying description {\it \`a la} Kerr/CFT that provides a microscopic foundation for the Bekenstein-Hawking  entropy of all these black holes by focusing only on a certain near-horizon region, its  CFT$_2$ and applying the Cardy formula was implemented in \cite{Nian:2020qsk, David:2020ems, David:2020jhp}.

\begin{figure}[!htb]
      \begin{center}
        \includegraphics[width=0.25\textwidth]{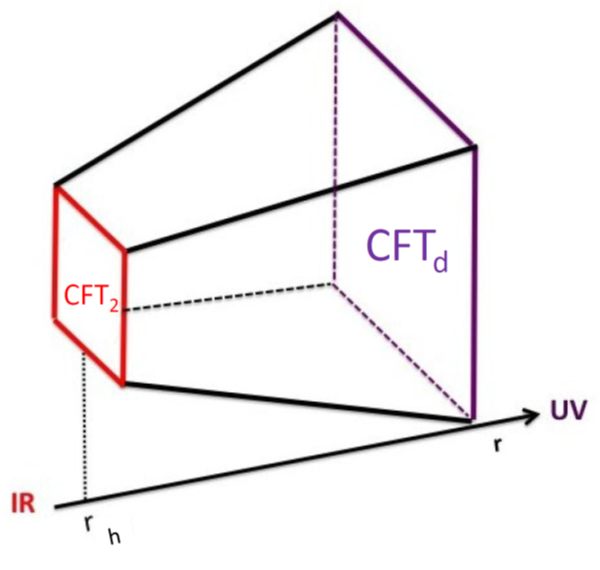}
      \caption{Renormalization group flow in a rotating AdS black hole}
      \label{fig:Holography Scale}
      \end{center}
    \end{figure}

The general situation is one in which for an asymptotically AdS$_{d+1}$ rotating black hole, we can apply the AdS$_{d+1}$/CFT$_d$ correspondence in the asymptotic boundary region and, concurrently, the Kerr/CFT correspondence in the near-horizon region. The near-horizon region has a local AdS$_3$ factor,  instead of the AdS$_2$ arising in previous studies \cite{Faulkner:2009wj, Faulkner:2010jy}, and correspondingly  insinuates a holographic renormalization group flow connecting a CFT$_d$ in the UV and a CFT$_2$ in the IR (Fig.~\ref{fig:Holography Scale}). The Kerr/CFT correspondence  works not only for AdS black holes but also for much wider classes of rotating black holes, however, the AdS black holes have the particularly nice feature of providing a UV completion via the dual conformal field theory as depicted in Fig.~\ref{fig:Holography Scale}.

In this manuscript we define a master equation for the retarded Green's function in the full background. Following  the canonical AdS/CFT dictionary into the asymptotic boundary we obtain a result that can be naturally interpreted in as a CFT$_3$ result. We also take an alternative new route and zoom in on the near-horizon region, blow up this region and show that it yields a  result for the retarded Green's function that fits in a  CFT$_2$ framework. The near-horizon approach is consistent because the particular near-horizon limit led to a background that, by itself, satisfies the equations of motion. This decoupling of the near-horizon region  is akin to the decoupling of the near-throat region of a D3-brane, which lead to the original formulation of the AdS/CFT correspondence. We, therefore, argue that in this context the Kerr/CFT correspondence follows as a decoupling of the standard AdS/CFT correspondence applied to rotating AdS black holes.

Having computed the  retarded Green's function, an interesting observable which we determine is  the shear viscosity in the background of near-extremal rotating AdS$_4$ black holes, generalizing the results first obtained in \cite{Policastro:2001yc} for black 3-branes which were further expanded and applied to  AdS black holes in \cite{Buchel:2003tz, Kovtun:2004de, Buchel:2004di, Kats:2007mq} with many subsequent following works.  We find that, due to the presence of  angular momentum, the shear viscosity to entropy density ratio is lower than the $1/(4\pi)$ bound, vanishes in a particular extremal limit, and grows quadratically for small temperatures, $\eta/s \sim T^2$.


{\bf Master Equation for Retarded Green's Function.---} 
The asymptotically AdS$_4$ Kerr black hole  is given by the metric
\begin{align}
  ds^2 & = - \frac{\Delta_r}{\rho^2} \left(dt - \frac{a\, \textrm{sin}^2 \theta}{\Xi} d\phi \right)^2 + \frac{\rho^2}{\Delta_r} dr^2 + \frac{\rho^2}{\Delta_\theta} d\theta^2 \nonumber\\
  {} & + \frac{\Delta_\theta}{\rho^2}\, \textrm{sin}^2 \theta\, \left(a\, dt - \frac{r^2 + a^2}{\Xi} d\phi \right)^2\, ,\label{eq:KerrAdS4BHmetric}\\
  \textrm{where} &\qquad \rho^2 \equiv r^2 + a^2\, \textrm{cos}^2 \theta\, ,\quad \Xi \equiv 1 - \frac{a^2}{L^2}\, ,\\
  \Delta_r & \equiv (r^2 + a^2) \left(1 + \frac{r^2}{L^2} \right) - 2 M r\, ,\quad \Delta_\theta \equiv 1 - \frac{a^2\, \textrm{cos}^2 \theta}{L^2}\, ,\nonumber
\end{align}
 where $a$ is the rotation parameter and $L$ denotes the AdS$_4$ radius.  The thermodynamic quantities, energy, angular momentum, temperature, entropy and angular velocity are given by
\begin{align}
\begin{split}
 E = \frac{M}{\Xi^2}\, ,\quad J = \frac{M a}{\Xi^2}\, , & \quad T = \frac{r_+ \Big[1 + \frac{a^2}{L^2} + \frac{3 r_+^2}{L^2} - \frac{a^2}{r_+^2} \Big]}{4 \pi (r_+^2 + a^2)}\, ,\\
 S = \frac{\pi (r_+^2 + a^2)}{\Xi}\, , & \quad \Omega = \frac{a \left(1 + \frac{r_+^2}{L^2} \right)}{r_+^2 + a^2}\, ,\label{eq:KerrAdS S_BH}
\end{split}
\end{align}
with $r_+$ denoting the outer horizon position, the largest root of $\Delta_r=0$. The extremal limit is achieved when $\Delta_r$ has a double root, which is equivalent to fixing $M$ as a function of $a$, {\it i.e. }, $M = M_{ext} (a)$ \cite{Caldarelli:1999xj}.

The first step toward the retarded Green's function is to consider the wave equation of an uncharged massless scalar in the background \eqref{eq:KerrAdS4BHmetric}
\be\label{eq:KerrWaveEq}
  \partial_\mu \left(\sqrt{- g}\, g^{\mu\nu} \partial_\nu \Psi \right) = 0\, .
\ee
 Here we consider a massless scalarto highlight certain properties. When compared to the standard AdS/CFT situation, we note that the typical translational symmetry along the directions of the brane is broken by in the context of the rotating black hole background. Such situation has been   considered recently in \cite{Garbiso:2020puw}; a previous relevant discussion was given in \cite{Sonner:2009fk}.  Breaking translational symmetry is quite relevant to address more realistic aspects of transport in condensed matter applications. The rotating black hole background breaks some translational symmetry in a natural way. There are, however, other ways of breaking the translational symmetry.  For instance, a particular class of models has been discussed in \cite{Andrade:2013gsa, Hartnoll:2016tri}, where the translational symmetry is  broken by certain scalars in the theory whose classical solution is linear in coordinates yielding an equation for the shear perturbation of the metric which behaves as a massive scalar equation.

To solve \eqref{eq:KerrWaveEq}, we assume an Ansatz of the form $\Psi = e^{- i \omega t + i m \phi}\, R(r)\, S (\theta)$. It is known that one can separate the wave equation \eqref{eq:KerrWaveEq} into its radial  and angular parts \cite{Carter:1968ks}
\begin{align}
  \frac{d}{dr} \bigg[\Delta_r \frac{d}{dr} R(r) \bigg] + V(r) \, R(r) & = 0 \, ,\label{eq:KerrWaveEqRadial}\\
  \frac{1}{\textrm{sin}\, \theta} \frac{d}{d\theta} \bigg[\Delta_\theta\, \textrm{sin}\, \theta\, \frac{d}{d\theta} S (\theta) \bigg] - \frac{(m \Xi)^2}{\Delta_\theta\, \textrm{sin}^2 \theta}\, S (\theta) \quad & {} \nonumber\\
  + \bigg[\frac{2 m \omega a \Xi}{\Delta_\theta} - \frac{a^2 \omega^2\, \textrm{sin}^2 \theta}{\Delta_\theta} \bigg] S (\theta) + K\, S (\theta) & = 0\, ,\label{eq:KerrWaveEqAngular}
\end{align}
where $K$ is the separation constant, and the potential $V(r)$ is
\be
  V (r) \equiv - K + \frac{\Big[\omega (r^2 + a^2) - a m \Xi \Big]^2}{\Delta_r (r)}\, .
\ee

\begin{figure}[!htb]
      \begin{center}
        \includegraphics[width=0.4\textwidth]{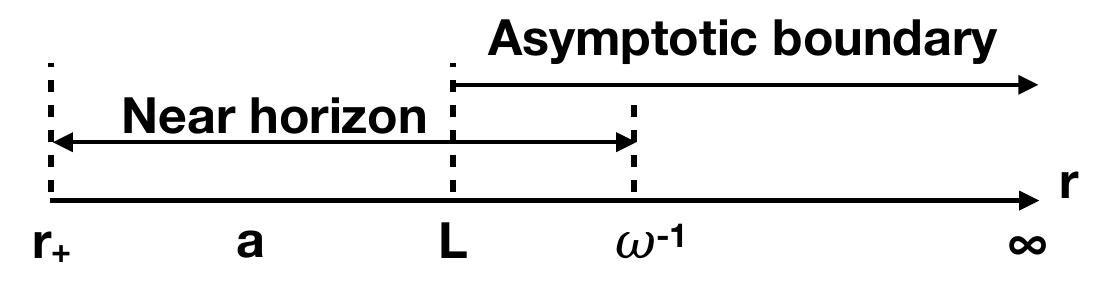}
      \caption{The near-horizon and the asymptotic boundary regions}
      \label{fig:Regions}
      \end{center}
    \end{figure}

The strategy for solving \eqref{eq:KerrWaveEq} is fairly standard. We consider  the wave equation \eqref{eq:KerrWaveEq} focusing on two regions: (a) the  asymptotic boundary region ($r \to \infty$) and (b) the near-horizon region ($\omega r \ll 1$). After solving in these regions  separately, we then glue the solutions in some overlapping region (see Fig.  \ref{fig:Regions} which contains more details). The canonical form of the AdS/CFT correspondence stipulates that from these solutions we can construct observables in the dual CFT$_3$ at the asymptotic boundary. Our new observation is that by zooming in on the near-horizon region we will also define observables in an effective  dual CFT$_2$. We argue that this process of zooming in is an explicit holographic realization of  a renormalization group connecting these conformal fixed points of different dimensions, namely, CFT$_3$ (defined in the asymptotic region) and CFT$_2$ (defined in the near-horizon region). 

Let us now discuss some technical details that are crucial to understand the regime of validity of our approximations.  As we have seen in Eqs.~\eqref{eq:KerrWaveEqRadial} and \eqref{eq:KerrWaveEqAngular}, the scalar wave equation \eqref{eq:KerrWaveEq} in the full AdS$_4$ black hole geometry factorizes into the radial part and the angular part as a consequence of the existence of Killing-Yano tensor \cite{Carter:1968ks}.  This separation allows us to focus on the radial part of equation and treat it  in the  standard  AdS/CFT prescription we would apply to a stationary black hole or the black D3 brane background. Of course, we need to recall that the separation constant $K$ is, in general a fairly complicated quantity that, in practice, needs to be determined numerically, as demonstrated in \cite{Sonner:2009fk}. We can, for simplicity, focus on states with quantum number $m=0$ above and with very low energies $\omega$ but should keep in mind that the general situation is quite more involved technically and requires a  numerical treatment.

Focusing on the radial part of the scalar Laplace equation \eqref{eq:KerrWaveEqRadial}  gives us a clue to understand the unification of the retarded Green's function for the boundary  CFT$_3$ and the near-horizon CFT$_2$ and how there is an interpolation between the two.  The solutions to the wave equation have a fairly generic behavior in the large-$r$ limits.  If we neglect all sub-leading terms, including of the form $a/r$, the solution in the asymptotic boundary region takes the form 
\be\label{eq:asympt sol near bdy}
  R (r) \stackrel{r \to \infty}{=} C\, r^{- \Delta_-} + D\, r^{\Delta_- - d} \quad\Rightarrow\quad G_R^{\textrm{CFT}_3} \sim \frac{D}{C}\, ,
\ee
with $d = 3$ and $\Delta_- = 0$ in this case; this is the expected behavior for a Green's function in a CFT$_3$.  In the near-horizon region, we define an effective radial coordinate $\widetilde{r}$ (see the precise definition in equation \eqref{eq:NH coord limit}) in terms of which the solution takes the form  
\be\label{eq:asympt sol near horizon}
  R (\widetilde{r}) \stackrel{\widetilde{r} \to \infty}{=} A\, \widetilde{r}\,^{- h} + B\, \widetilde{r}\,^{h - 1} \quad\Rightarrow\quad G_R^{\textrm{CFT}_{2}} \sim \frac{B}{A}\, ,
\ee
where $h  \to 0$ in the low-energy limit ($\omega \to 0$, $m \to 0$),  signaling a result in a CFT$_2$.

 It is important to verify that the near-horizon geometry satisfies the equations of motion on its own. This seemingly banal statement is quite crucial, note, for example,  that for generic  non-extremal configurations the limit might not even exist as a smooth geometry due to the divergences that remain when explicitly applying the limiting procedure dictated  by the equation \eqref{eq:NH coord limit}.   When the near-horizon geometry is smooth and independently solves the equations of motion, we can then view that geometry and its field theory dual as a stand-alone correspondence. The prototypical example is the D3-brane and its near-horizon geometry AdS$_5\times S^5$, where the latter leads to a stand-alone correspondence that needs not refer any more to the starting D3-brane geometry. Here too, having a near-horizon geometry that solves the equations of motion on its own allows us to view the Kerr/CFT$_2$ as a stand-alone conjecture. Of course, from the RG flow point of view, the Kerr/CFT$_2$ is the IR of a well-defined AdS$_4$/CFT$_3$ setup for rotating black holes.

To implement the near-horizon limit, it is convenient that we can work in truncated gauged supergravity theories. For example, one can view the background \eqref{eq:KerrAdS4BHmetric} as a solution of 4d minimal gauged supergravity. More generally, this class of solutions can be uplifted to 10d or 11d. In fact, there are different ways of truncating the 10d or 11d gauged supergravity theories to lower dimensions, including the ones without dilatons (see \cite{David:2020ems} and the references therein) and the ones with dilatons \cite{Cvetic:1999xp, Liu:1999ai, Chamseddine:1999xk}. In the class of solutions with dilatons, \cite{Gubser:2009qt} has found a near-horizon AdS$_3$ from a charged dilatonic AdS$_5$ black hole. Although the near-horizon AdS$_3$ in \cite{Gubser:2009qt} was not used to perform a AdS$_3$/CFT$_2$ similar to the Kerr/CFT, it was emphasized in \cite{Gubser:2009qt} that the near-horizon solution uplifted to 10d is indeed by itself a solution to the equations of motion of IIB supergravity.

In general, for a  solution  $R(r)$  to a differential equation $R''(r) + P\, R'(r) + Q\, R(r) = 0$, we can prove a {\it Master Equation}  for the retarded Green's function \cite{CaronHuot:2006te, Atmaja:2008mt}
\be\label{eq:Green fct in Wronskian}
  \textrm{Im}\, G_R^{\textrm{CFT}} \sim \lim_{r \to \infty} r^{-2 \tilde{\Delta}} \frac{W (r)}{2 i R^* (r)\, R (r)} \, ,
\ee
where $\tilde{\Delta}$ can be $\Delta_-$ or $h$ above, and we used  the  Wronskian 
\be\label{eq:def Wronskian}
  W (r) \equiv e^{\int^r P(r') dr'}\, \left[R^* (r)\, \partial_r R (r) - R (r)\, \partial_r R^* (r) \right]\, .
\ee
Eqs.~\eqref{eq:Green fct in Wronskian} and \eqref{eq:def Wronskian} apply to both the asymptotic boundary CFT$_3$ and the near-horizon CFT$_2$. When applied to the near-horizon CFT$_2$, the $r$ in \eqref{eq:Green fct in Wronskian} and \eqref{eq:def Wronskian} should be replaced by $\widetilde{r}$ which starts its life as a near-horizon coordinate but it is later assumed to run as an infinite affine parameter.  Given incoming boundary conditions at the horizon, the {\it Master Equation} \eqref{eq:Green fct in Wronskian} governs the retarded Green's function along the whole RG flow, because the Wroskian $W (r)$ is conserved along $r$, i.e. $\partial_r W(r) = 0$. We argue that focusing on different regions of the geometry is akin to implementing a Kadanoff style coarse-graining. Namely,  following  the canonical AdS/CFT dictionary and applying \eqref{eq:Green fct in Wronskian} we will get a CFT$_3$ result. We also take a new route and zoom in on the near-horizon region, blow up this region out and show that  in this case the {\it Master Equation} \eqref{eq:Green fct in Wronskian} yields a CFT$_2$ result.

{\bf CFT$_3$ Retarded Green's Function from Asymptotic Boundary.---} 
 In the asymptotic region corresponding to  $r \to \infty$ ($r$ larger than any other scale in the problem) near the asymptotic conformal boundary, we keep only terms of order up to $\mathcal{O} (r^{-4})$ in the radial part of the wave equation \eqref{eq:KerrWaveEq}, i.e. \eqref{eq:KerrWaveEqRadial}, which can be expressed as
\be\label{eq:KerrWaveEqRadial Near Boundary}
  R''(r) + \left[\frac{4}{r} - \frac{2 (a^2 + L^2)}{r^3} \right] R'(r) + \frac{\omega^2 L^4 - K L^2}{r^4} R(r) = 0\, .
\ee
This differential equation has the following analytical solution
\begin{align}
  {} & R(r) = C_1\cdot \phantom{|}_1 F_1 \left(\frac{\omega^2 L^4 - K L^2}{4 (a^2 + L^2)},\, -\frac{1}{2};\, - \frac{a^2 + L^2}{r^2} \right) \label{eq:Exact Sol Near Boundary Kerr}\\
  {} & \quad + C_2\, \frac{(a^2 + L^2)^{\frac{3}{2}}}{r^3}\cdot \phantom{|}_1 F_1 \left(\frac{3}{2} + \frac{\omega^2 L^4 - K L^2}{4 (a^2 + L^2)},\, \frac{5}{2};\, - \frac{a^2 + L^2}{r^2} \right)\, ,\nonumber
\end{align}
with the  constants $C_{1, 2}$ determined by gluing the asymptotic boundary and the near-horizon solutions, which will be discussed in the next section.

Let us be more careful in explaining the regime in which we detect a CFT$_3$ behavior. The exact analytic solution \eqref{eq:Exact Sol Near Boundary Kerr} contains various terms. We can roughly track their origins direction in the equation \eqref{eq:KerrWaveEqRadial Near Boundary}. Namely,  there are terms of the form: $\frac{1}{r}, \frac{a^2}{r^3}, \frac{L^2}{r^3}$ as well as those proportional to $\frac{1}{r^4}$ that we have kept.  We can further consider the low-energy limit ($\omega \to 0$, $K \to 0$) for which the first term  in the analytic solution \eqref{eq:Exact Sol Near Boundary Kerr} has the asymptotic  (as $r\to \infty$) behavior $1 + \mathcal{O} (r^{-2})$ while the second term goes as $\frac{(a^2 + L^2)^{\frac{3}{2}}}{r^3} + \mathcal{O} (r^{-5})$. This behavior, 
 together with the phase factor corresponds to the wave $e^{- i \omega t + i m \phi}\, S (\theta)\, (a^2 + L^2)^{\frac{3}{2}} / r^3$ for $r \to \infty$. Hence, the second term in \eqref{eq:Exact Sol Near Boundary Kerr} can be seen to correspond to an  incoming wave due to its negative phase velocity. Neglecting further the effects of rotation (factors proportional to $a$) in the {\it far UV region}, basically neglecting, contributions of the form $\sim \frac{a^2 L}{r^3}$, the wave  equation has the asymptotic solution near the boundary
\be
R (r) \stackrel{r \to \infty}{=}  C_1\cdot r^{\Delta_+ - d} = C_1\cdot r^{- \Delta_-}\, ,
\ee
where $\Delta_\pm = \frac{d}{2} \pm \frac{1}{2} \sqrt{d^2 + 4 \mu^2 L^2}$ for a scalar with mass $\mu$ in AdS$_{d+1}$ space. For the  massless scalar ($\mu$ = 0) close to the boundary of the AdS$_4$ black hole there should be
\be
  \Delta_+ = 3\, ,\quad \Delta_- = 0\, ,\quad d = 3\, ,
\ee
which exactly matches the asymptotic behaviors of the two terms near the boundary ($r \to \infty$)  in \eqref{eq:Exact Sol Near Boundary Kerr}. This result is at the core of the AdS/CFT correspondence \cite{Gubser:1998bc, Witten:1998qj}, particularly for a boundary CFT$_3$ since  the scalar in the bulk will couple to a boundary operator with the conformal dimension $\Delta_+ = 3$ and the 2-point spinless correlator $\langle \mathcal{O} (x)\, \mathcal{O} (y) \rangle \sim 1 / |x - y|^{2 \Delta_+}$. We need to be, of course, a bit more careful with this analysis, as we have neglected all traces of the rotation, and no factors of $a/r$ have been kept in this limit. We could, restore that dependence in a way that will imply a departure from the correlator we have written, which naturally corresponds to a field theory defined on $\mathbb{R}^3$, rather than on $S^1\times S^2$ as we have been discussing.

In principle, this expression is only valid at the UV conformal fixed point, and we cannot simply read off the IR CFT$_2$ correlator from the solution near the asymptotic boundary. To obtain the information of the IR CFT$_2$, we need to analyze the solution in the near-horizon region of the AdS black hole.  As discussed before, we glue the asymptotic boundary and the near-horizon solutions at $r \to \infty$ to fix the constant $C_1$ in \eqref{eq:Exact Sol Near Boundary Kerr}. We can do it in a more precise way by gluing them at some intermediate $r_B$ in the overlapping region with $L \ll r_B \ll \omega^{-1}$. Consequently, the retarded Green's function of the boundary CFT$_3$ is affected  by the renormalization scale $r_B$, {\it i.e.},  $G_R (\omega, m, K; r_B)$. In the low-energy limit, $G_R (r_B) \sim 1 + \mathcal{O} (r_B^{-4})$, which trivially satisfies the Callan-Symanzik equation $M \frac{\partial}{\partial M} G_R (r_B) = 0$ with $M \sim r_B / L^2$, implying the vanishing of the anomalous dimension of a certain dimension-3 operator \cite{Gubser:1998bc}.

{\bf Near-Horizon CFT$_2$ Retarded Green's Function.---} In the near-horizon region ($\omega r \ll 1$), we can expand $\Delta_r (r)$ to quadratic order in $r - r_+$
\begin{align}
  \Delta_r (r) & = k (r - r_+) (r - r_*) + \mathcal{O} \left((r - r_+)^3 \right)\, ,\\
  \textrm{where}\qquad k & = 1 + \frac{a^2}{L^2} + \frac{6 r_+^2}{L^2}\, ,\\
  r_* & = r_+ - \frac{1}{k r_+} \left(r_+^2 - a^2 + \frac{3 r_+^4}{L^2} + \frac{a^2 r_+^2}{L^2} \right)\, .\nonumber
\end{align}
To zoom in on the near-horizon region we follow a limiting procedure first discussed for asymptotically flat spacetimes in \cite{Bardeen:1999px} and implemented for asymptotically AdS ones in \cite{Chen:2010bh}
\be\label{eq:NH coord limit}
  r \to \frac{r_+ + r_*}{2} + \epsilon r_0 \widetilde{r}\, ,\quad t \to \widetilde{t}\, \frac{r_0}{\epsilon}\, ,\quad \phi \to \widetilde{\phi} + \Omega_H \widetilde{t}\, \frac{r_0}{\epsilon}\, ,
\ee
where $\Omega_H \equiv \frac{a \Xi}{r_+^2 + a^2}$, $r_+ - r_* = \lambda \epsilon r_0$,  $r_0^2 \equiv \frac{r_+^2 + a^2}{k}$, and $\epsilon$ is a small positive scaling parameter, which will be eventually sent to zero,  while the parameter $\lambda$ characterizes how far the near-extremal black hole is from extremality.  In the limit $\epsilon \to 0$, we obtain the near-horizon metric for the near-extremal AdS$_4$ Kerr black hole \cite{Chen:2010bh}:
\begin{align}\label{eq:Kerr NH metric}
  ds^2 & = \Gamma (\theta) \Bigg[- \left(\widetilde{r} - \frac{\lambda}{2} \right) \left(\widetilde{r} + \frac{\lambda}{2} \right) d\widetilde{t}^2 + \frac{d\widetilde{r}^2}{\left(\widetilde{r} - \frac{\lambda}{2} \right) \left(\widetilde{r} + \frac{\lambda}{2} \right)} + \alpha (\theta)\, d\theta^2 \Bigg] \nonumber\\
  {} & + \gamma (\theta) \left[d\widetilde{\phi} + \widetilde{p}\, \left(\widetilde{r} - \frac{\lambda}{2}\right) d\widetilde{t}\right]^2\, ,
\end{align}
with the corresponding factors
\begin{align}
   \Gamma (\theta) = \frac{\rho_+^2\, r_0^2}{r_+^2 + a^2}\, ,\quad \alpha (\theta) = \frac{r_+^2 + a^2}{\Delta_\theta\, r_0^2}\, ,\quad & \gamma (\theta) = \frac{\Delta_\theta\, \textrm{sin}^2 \theta\, (r_+^2 + a^2)^2}{\rho_+^2\, \Xi^2}\, ,\nonumber\\
   \widetilde{p} = \frac{a\, r_0^2\, \Xi (r_+ + r_*)}{(r_+^2 + a^2)^2}\, ,\quad \rho_+^2 = & r_+^2 + a^2\, \textrm{cos}^2 \theta\, .\label{eq:Kerr NH metric factors}
\end{align} 
The goal is to study the wave equation \eqref{eq:KerrWaveEq} in the background \eqref{eq:Kerr NH metric} obtained by the limit $\epsilon \to 0$ above. This limiting procedure on the gravity side resembles Kadanoff's original block spin coarse-graining.

We implement the following Ansatz for the solution to the wave equation. Note that the frequencies are scaled to the superradiant bound \cite{Bredberg:2009pv}
\be
  \Psi = e^{- i \hat{\omega} \widetilde{t} + i m \widetilde{\phi}}\, R (\widetilde{r})\, S (\theta)\, ,\quad \textrm{with}\quad \hat{\omega}\, \frac{\epsilon}{r_0} \equiv \omega - m \Omega_H\, .
\ee
 A Kerr black hole has superradiant instabilities, which for a massless scalar happen when $\omega < m \Omega_H$ (see e.g. \cite{Cardoso:2004nk}). Therefore, as long as we consider $\hat{\omega} \geq 0$, our near-horizon solution is free of superradiant instability.

For the radial part of the wave equation, we can solve an alternative equation, which in the near-horizon limit ($\hat{\omega} r \ll 1$, $r \geq r_+$) and close to the superradiant bound ($\hat{\omega} a \ll 1$) is equivalent to the origional radial equation \eqref{eq:KerrWaveEqRadial}  with the near-horizon scaling \eqref{eq:NH coord limit} at the leading order
\be\label{eq:KerrWaveEqRadialNew}
  \frac{d}{d\widetilde{r}} \left[ \left(\widetilde{r}^2 - \frac{\lambda^2}{4} \right) \frac{dR(\widetilde{r})}{d\widetilde{r}} \right] + \frac{\lambda}{\widetilde{r} - \frac{\lambda}{2}} \hat{A}\, R (\widetilde{r}) + \frac{\lambda}{\widetilde{r} + \frac{\lambda}{2}} \hat{B}\, R (\widetilde{r}) + \hat{C}\, R (\widetilde{r}) = 0\, ,
\ee
\be
  \textrm{where}\qquad \hat{A} \equiv \frac{\hat{\omega}^2}{\lambda^2}\, ,\quad \hat{B} \equiv - \left(\frac{\hat{\omega}}{\lambda} - \frac{2 r_+\, m\, \Omega_H}{k} \right)^2\, ,\quad \hat{C} \equiv - \frac{\hat{K}}{k}\, ,\nonumber
\ee
and $\hat{K}$ is a separation constant. This equation can also be obtained directly in the near-horizon geometry \eqref{eq:Kerr NH metric} instead of taking the near-horizon limit of the full wave equation \eqref{eq:KerrWaveEqRadial}, and these two derivations lead the same equation \eqref{eq:KerrWaveEqRadialNew} in the limits $\hat{\omega} r \ll 1$ and $\hat{\omega} a \ll 1$.

In the coordinate $\widetilde{z} \equiv \frac{\widetilde{r} - \lambda / 2}{\widetilde{r} + \lambda / 2}$, the radial wave equation \eqref{eq:KerrWaveEqRadialNew} in the near-horizon region has an exact solution
\begin{align}
  R (\widetilde{z}) = \widetilde{z}^{\hat{\alpha}} (1 & - \widetilde{z})^{\hat{\beta}}\, \phantom{|}_2 F_1(\hat{a},\, \hat{b},\, \hat{c};\, \widetilde{z})\, ,\label{eq:Exact Sol Near Horizon Kerr}\\
  \textrm{with}\qquad \hat{\alpha} \equiv - i \sqrt{\hat{A}}\, , & \quad \hat{\beta} \equiv \frac{1}{2} \left(1 - \sqrt{1 - 4 \hat{C}} \right)\, ,\\
  \hat{a} \equiv \hat{\alpha} + \hat{\beta} + i \sqrt{- \hat{B}}\, , & \quad \hat{b} \equiv \hat{\alpha} + \hat{\beta} - i \sqrt{- \hat{B}}\, ,\quad \hat{c} \equiv 1 + 2 \hat{\alpha}\, .\nonumber
\end{align}
In the region very close to the horizon ($\widetilde{z} \to 0$), the asymptotic behavior of $R(\widetilde{z})$ is
\be
  R (\widetilde{z}) \stackrel{\widetilde{z} \to 0}{=} \widetilde{z}^{\hat{\alpha}} (1 + \mathcal{O} (\widetilde{z})) \quad\sim\quad \left[\frac{1}{\lambda} \left(\widetilde{r} - \frac{\lambda}{2} \right) \right]^{-i \frac{\hat{\omega}}{\lambda}}\, .
\ee
Hence, the wave solution $\Psi = e^{- i \hat{\omega} \widetilde{t} + i m \widetilde{\phi}}\, R (\widetilde{r})\, S (\theta)$ can be interpreted as an infalling wave at the horizon due to its negative phase velocity.

To obtain the asymptotic behavior of $R(\widetilde{z})$ at $\widetilde{r} \to \infty$ or equivalently $\widetilde{z} \to 1$ in the near-horizon region, we use some properties of the hypergeometric function $\phantom{|}_2 F_1 (a, b, c; z)$ to rewrite the solution \eqref{eq:Exact Sol Near Horizon Kerr}.
In the coordinate $\widetilde{r}$, the asymptotic behavior of $R(\widetilde{r})$ for $\widetilde{r} \to \infty$ is
\begin{align}\label{eq:KerrWaveEqRadial AsympSol}
  R (\widetilde{r}) \stackrel{\widetilde{r} \to \infty}{=} A\, & \widetilde{r}^{\hat{h} - 1} + B\, \widetilde{r}^{- \hat{h}}\, ,\\
  \textrm{where}\qquad\qquad \hat{h} \equiv \frac{1}{2} \Big(1 + & \sqrt{1 - 4 \hat{C}} \Big)\, ,\\
  A \equiv \lambda^{1 - \hat{h}}\, \frac{\Gamma (\hat{c})\, \Gamma (\hat{c} - \hat{a} - \hat{b})}{\Gamma (\hat{c} - \hat{a})\, \Gamma (\hat{c} - \hat{b})} \, , & \quad B \equiv \lambda^{\hat{h}}\, \frac{\Gamma (\hat{c})\, \Gamma (\hat{a} + \hat{b} - \hat{c})}{\Gamma (\hat{a})\, \Gamma (\hat{b})}\, .\nonumber
\end{align}
Due to the condition $\omega L \ll 1$, which is a consequence of $r_+ / L \ll 1$ and $\omega T \ll 1$, the asymptotic boundary region ($r \gg L$) and the near-horizon region ($\omega r \ll 1$) have some overlapping region ($L \ll r \ll \omega^{-1}$). Hence, we can glue the asymptotic solution \eqref{eq:Exact Sol Near Boundary Kerr} with the near-horizon solution \eqref{eq:Exact Sol Near Horizon Kerr} in the overlapping region $L \ll r \ll \omega^{-1}$ (see Fig.~\ref{fig:Regions}).
    \\

We follow the approach in \cite{Policastro:2001yc} by comparing the $r \to \infty$ behavior \eqref{eq:KerrWaveEqRadial AsympSol} of the solution in the near-horizon region:
\be
  R (r) \stackrel{r \to \infty}{\to} 1 \quad \textrm{for}\quad \hat{\omega}, m, \hat{K} \to 0
\ee
with the $r \to \infty$ behavior \eqref{eq:Exact Sol Near Boundary Kerr} of the solution in the asymptotic boundary region:
\be
  R (r) \stackrel{r \to \infty}{\to} C_1 \quad \textrm{for}\quad \omega, K \to 0\, ,
\ee
which fixes the constant $C_1$ to be
\be
  C_1 = 1\, .
\ee
The explicit value of $C_2$ can be fixed by requring that the ratio of the shear viscosity and the entropy density approaches the known value $1/(4 \pi)$ in the non-rotating limit $a \to 0$. The result is
\be
  C_2 = \frac{24 \sqrt{2}}{\pi (\omega L)^{\frac{7}{2}}}\, .
\ee
Although the value of $C_2$ is large due to $\omega L \ll 1$, for $r \to \infty$ the first term in \eqref{eq:Exact Sol Near Boundary Kerr} of order $\mathcal{O} (r^0)$ is still the dominant one compared to the second term of order $\mathcal{O} (r^{-3})$.

With the following identifications of parameters
\begin{align}
\begin{split}\label{eq:Kerr AdS near horizon parameter identification}
  \hat{T}_L & \equiv \frac{k}{4 \pi r_+ \Omega_H}\, ,\quad \hat{T}_R \equiv \frac{k r_0}{4 \pi a \Xi} \lambda \epsilon\, ,\\
  \hat{\omega}_L & \equiv m\, ,\quad \hat{\omega}_R \equiv \frac{r_0 k}{a \Xi} \left(\hat{\omega} - \frac{\lambda r_+ m \Omega_H}{k} \right) \epsilon\, ,
\end{split}
\end{align}
the retarded Green's function can be computed
\be\label{eq:KerrGreenFct}
  G_R = \frac{B}{A} = \lambda^{2 \hat{h} - 1}\, \frac{\Gamma (1 - 2 \hat{h})}{\Gamma (2 \hat{h} - 1)} \frac{\Gamma \left(\hat{h} - i\, \frac{\hat{\omega}_L}{2 \pi \hat{T}_L} \right)\, \Gamma \left(\hat{h} - i\, \frac{\hat{\omega}_R}{2 \pi \hat{T}_R} \right)}{\Gamma \left(1 - \hat{h} - i\, \frac{\hat{\omega}_L}{2 \pi \hat{T}_L} \right)\, \Gamma \left(1 - \hat{h} - i\, \frac{\hat{\omega}_R}{2 \pi \hat{T}_R} \right)}\, .
\ee
Its imaginary part exactly matches the absorption cross section in the near-horizon region, and it is exactly the CFT$_2$ retarded two-point function of the CFT$_2$ operator with $h_L = h_R = \hat{h}$ dual to the scalar field \cite{Bredberg:2009pv, Chen:2010bh}.

{\bf Shear Viscosity from AdS$_4$ Kerr Black Hole.---} Let us now turn to a particular property of the strongly coupled boundary CFT$_3$  dual  to the rotating asymptotically AdS$_4$ black hole -- the shear viscosity to entropy density ratio --  by  following the pioneering work \cite{Policastro:2001yc}.  More precisely, the authors of \cite{Policastro:2001yc} originally considered the black brane background, similar analysis were later applied to more general AdS black holes in \cite{Buchel:2003tz, Kovtun:2004de}.

For rotating AdS$_4$ black holes, translational symmetry is broken and, consequently, the shear viscosity has only one independent component, which is the counterpart of $\eta_{xyxy}$ for the AdS$_5$ case defined in \cite{Rebhan:2011vd} 
\begin{align}
  \eta_{xyxy} & = - \lim_{\omega \to 0} \frac{1}{\omega}\, \textrm{Im}\, G^R_{T^{xy} T^{xy}} (\omega, k=0) \nonumber\\
  {} & = \lim_{\omega \to 0} \frac{1}{2 \omega} \int dt\, d\vec{x}\, e^{i \omega t} \langle \big[T^{xy} (t, \vec{x}),\, T^{xy} (0, 0) \big] \rangle\, .
\end{align}
The boundary geometry of an AdS$_4$ Kerr black hole is the time circle together with a rotating 2-sphere, this is obtained by taking $r\to \infty$ in the metric  \eqref{eq:KerrAdS4BHmetric}. The resulting background breaks translational symmetry but still has the rotational symmetry. We denote the rotation axis of $S^2$ as the $z$-direction, while in the limit $r \to \infty$ the boundary $S^2$ locally can be viewed as flat $\mathbb{R}^2$, hence, we use the coordinates $x$ and $y$ on $\mathbb{R}^2$ to denote the directions perpendicular to the rotation axis, i.e. the $z$-direction, and the corresponding component of the energy momentum tensor is denoted by $T^{xy}$. Alternatively, the coordinates $x$ and $y$ can also be obtained from the conformal mapping of $S^2$ to $\mathbb{R}^2$ via a stereographic projection. Although the standard fluid/gravity correspondence on curved backgrounds was formulated for stationary backgrounds \cite{Bhattacharyya:2008mz}, small fluctuations around the equilibrium have also been considered when keeping the first dissipative gradient terms (see e.g. \cite{Bhattacharyya:2007vs}). Hence, the energy momentum tensor of the CFT on the boundary of an AdS$_4$ Kerr black hole has been shown to  still obey the hydrodynamics equations.

The fact that the  energy momentum tensor satisfies the hydrodynamics equations allows us to turn to the Kubo formula, according to which 
 the shear viscosity is related to the correlator of the stress tensor as \cite{Policastro:2001yc, Hartnoll:2016tri}
\begin{align}
  \eta & = \lim_{\omega \to 0} \frac{1}{2 \omega} \int dt\, d\vec{x}\, e^{i \omega t}\, \langle \left[ T_{xy} (t, \vec{x}),\, T_{xy} (0, 0) \right] \rangle \nonumber\\
  {} & = \lim_{\omega \to 0} \frac{1}{2 \omega i} \left[G_A (\omega) - G_R (\omega) \right]\, .
\end{align}
Similar to the often-discussed AdS$_5$ black hole case, it is more convenient to consider the AdS$_4$ black hole uplifted to 11d. More preciely, we consider in the following an 11d space  asymptoting to AdS$_4 \times$ $S^7$, which can be viewed as the near-horizon limit of M2-branes. Based on the AdS/CFT correspondence the absorption cross section $\sigma (\omega)$ of a graviton with frequency $\omega$ is also related to the correlator of the dual field theory
\be
  \sigma (\omega) = \frac{\kappa_{11}^2}{\omega} \int dt\, d\vec{x}\, e^{i \omega t}\, \langle \left[ T_{xy} (t, \vec{x})\, , T_{xy} (0, 0) \right] \rangle\, .
\ee
Therefore,
\be\label{eq:etaANDsigma0}
  \eta = \frac{1}{2 \kappa_{11}^2} \sigma (0)\, ,
\ee
where $\kappa_{11}$ is related to the 11-dimensional Newton's constant by $\kappa_{11} = \sqrt{8 \pi G_{11}}$. In asymptotically AdS$_4\times S^7$ solutions we further have $N \sqrt{2}\, (2 \pi^2 \kappa_{11})^{1/3} = 6 \Omega_7 L^6 / (\sqrt{2}\, \kappa_{11})$ \cite{Klebanov:1997kc}.

 Thus, equation \eqref{eq:etaANDsigma0} allows us to compute the viscosity by computing the 
absorption cross section $\sigma$. According to the AdS/CFT dictionary, we  interpret the two terms in the asymptotic boundary solution \eqref{eq:Exact Sol Near Boundary Kerr} as the background (source) and the incoming wave (response) respectively. Following \cite{Policastro:2001yc}, we first compute the absorption probability $P$ given by the ratio of the flux at $r = r_+$ and the flux from the incoming wave near the boundary $r = L$
\be\label{eq:Probability}
  P =  \frac{|\Psi (r = r_+)|^2 \cdot \textrm{area} (r = r_+)}{|\Psi (r = L)|^2\cdot \textrm{area} (r = L)}  = \frac{\pi^2 \omega^7 r_+^2 L^{11}}{1152\, (a^2 + L^2)^3}\, .
\ee
Let us remark that in \cite{Policastro:2001yc}, Eq.~\eqref{eq:Probability} was applied to  static black branes. For the near-extremal AdS$_4$ Kerr black hole considered in this paper, this relation still holds in the presence of rotation, because, for sufficiently low energies, the wave equation in this background completely factorizes into the radial part and the angular part, as we can see from Eqs.~\eqref{eq:KerrWaveEqRadial} and \eqref{eq:KerrWaveEqAngular}. The absorption cross section $\sigma$ is then related to the absorption probability $P$ for AdS$_4 \times$ $S^7$ by \cite{Das:1996we}
\be\label{eq:sigma0}
   \sigma (0) = \frac{384 \pi^3 P}{\omega^7} = \frac{\pi^5 r_+^2 L^{11}}{3 (a^2 + L^2)^3}\, .
\ee

\begin{figure}[!hbt]
\begin{center}
  \includegraphics[width=0.38\textwidth]{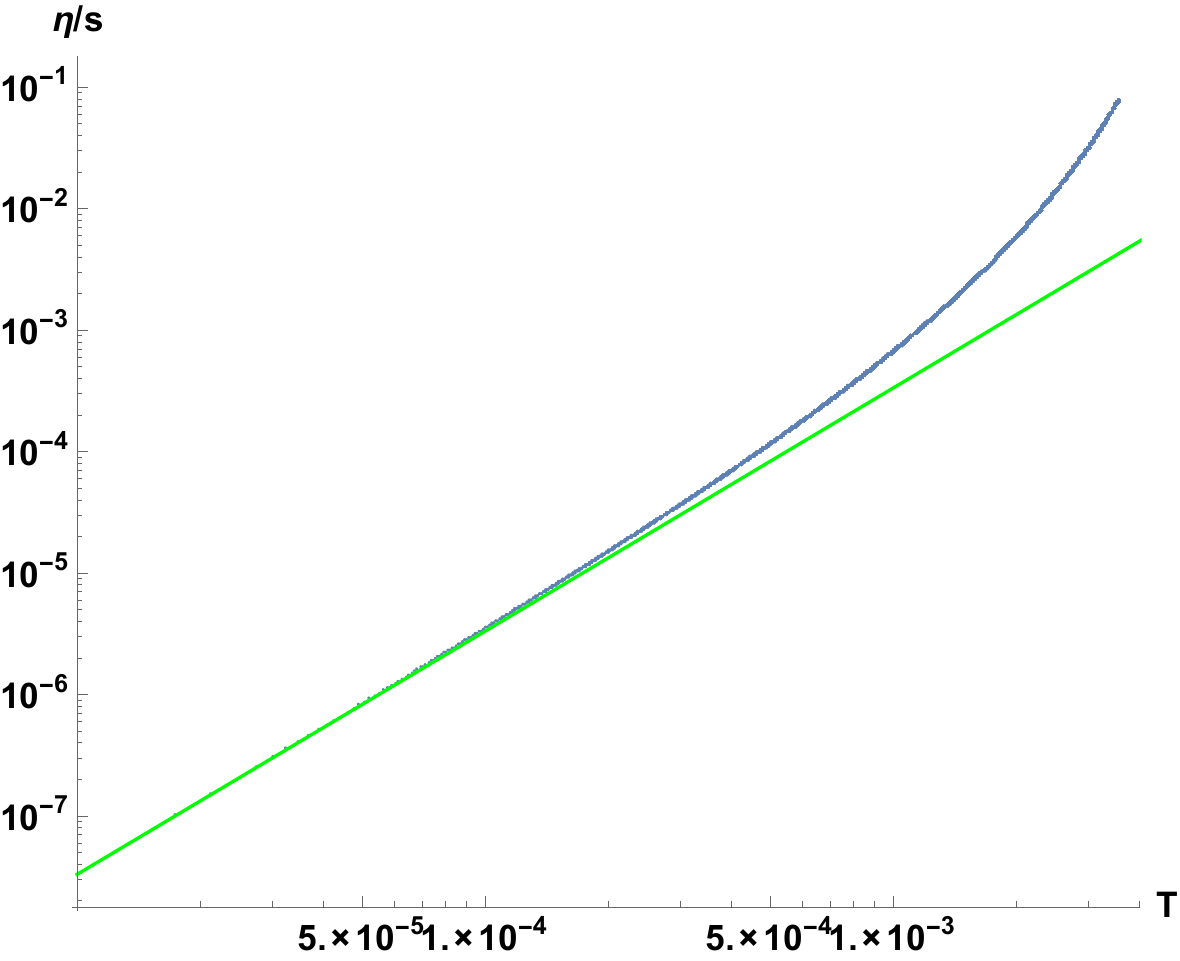}
  \caption{Near-extremal Kerr AdS$_4$ black hole $\eta / s$ vs $T$ for $L = 100$, with $\eta / s \sim T^2$ for small temperatures}
  \label{fig:Kerr BH eta over s VS T}
\end{center}
\end{figure}

Combining Eq.~\eqref{eq:etaANDsigma0} and Eq.~\eqref{eq:sigma0}, we obtain the shear viscosity $\eta$
\be
  \eta = \frac{N^{\frac{3}{2}} r_+^2 L^2}{3 \sqrt{2} (a^2 + L^2)^3}\, .
\ee
We can restore the Newton's constant $G_4 = \frac{3 L^2}{2 \sqrt{2} N^{3/2}}$ and compute the entropy density
\be
  s = \frac{2 \sqrt{2} N^{\frac{3}{2}} \pi (a^2 + r_+^2)}{3 L^2 (L^2 - a^2)}\, .
\ee 
Collecting all the previous partial results,
 we obtain
\be\label{eq: eta s ratio Kerr BH}
  \frac{\eta}{s} = \frac{r_+^2 L^4 (L^2 - a^2)}{4 \pi (a^2 + r_+^2) (a^2 + L^2)^3}\, .
\ee
The result \eqref{eq: eta s ratio Kerr BH} shares some features with the AdS$_5$ case studied recently in \cite{Garbiso:2020puw}. A few remarks are in order. First, as expected, the value of $\eta / s$ approaches the bound $\frac{1}{4 \pi}$ in the non-rotating limit $a \to 0$,  which has been computed in \cite{Kats:2007mq} for non-rotating AdS$_D$ ($D \geq 3$). Second, the viscosity to entropy density ratio, $\eta / s$, decreases from the value $1 / (4 \pi)$  as the black hole angular momentum ($\sim a$) increases, consistent with the result from breaking translational symmetry \cite{Burikham:2016roo}. For near-extremal AdS$_4$ Kerr black holes, this phenomenon can be understood in terms of the temperature $T$, which also decreases as $a$ increases.  In Fig.~\ref{fig:Kerr BH eta over s VS T} we indicate how at  very small temperatures we find numerically $\eta / s \sim T^2$.  This result precisely agrees with  other approaches to breaking translational symmetry such as those of \cite{Jain:2014vka, Jain:2015txa, Hartnoll:2016tri},  and follows from the scaling properties of the near-extremal, near-horizon geometry \cite{Hartnoll:2016tri}. The reason is that when the near-extremal, near-horizon geometry has an emergent IR scaling symmetry, the most relevant scale is the temperature, and consequently following the analysis of entropy production in \cite{Hartnoll:2016tri} we can conclude that the low-temperature behavior is $\eta / s \sim T^2$, which is indeed the case in this paper. More interesting, for extremal AdS$_4$ Kerr black holes $\eta / s$ vanishes when $a \to L$, because the entropy density blows up in this limit \cite{Hartnoll:2016tri}.

{\bf AdS$_4$ Kerr-Newman Black Hole.---} For potential condensed matter applications it is instructive to introduce chemical potential by considering the asymptotically AdS$_4$ Kerr-Newman black hole, which is formally  given by the same metric \eqref{eq:KerrAdS4BHmetric} as the Kerr black hole but with a different factor 
\be
  \Delta_r \equiv (r^2 + a^2) \left(1 + \frac{r^2}{L^2} \right) - 2 M r + q^2\,\,\, \textrm{with}\,\,\, q^2 \equiv q_e^2 + q_m^2\, ,
\ee
and an associated  Maxwell field
\be
  A = - \frac{q_e\, r}{\rho^2} \left(dt - \frac{a\, \textrm{sin}^2 \theta}{\Xi} d\phi \right) - \frac{q_m\, \textrm{cos}\, \theta}{\rho^2} \left(a\, dt - \frac{r^2 + a^2}{\Xi} d\phi \right)\, .
\ee
We focus on the electric case $q = q_e,\, q_m = 0$, for which the corresponding chemical potential is  read off as the difference
 \be
  \Phi = A_\mu \xi^\mu \Big|_{r \to \infty} - A_\mu \xi^\mu \Big|_{r = r_+} = \frac{q_e r_+}{r_+^2 + a^2}\, ,
\ee
where $\xi = \partial_t + \Omega_H\, \partial_\phi$ with $\Omega_H = \frac{a \Xi}{r_+^2 + a^2}$. The extremal limit is achieved when $\Delta_r$ has a double root, which is equivalent to fixing $M$ as a function of $a$ and $q$, {\it i.e.}, $M = M_{ext} (a, q)$ \cite{Caldarelli:1999xj}.

\begin{figure}[!htb]
      \begin{center}
        \includegraphics[width=0.45\textwidth]{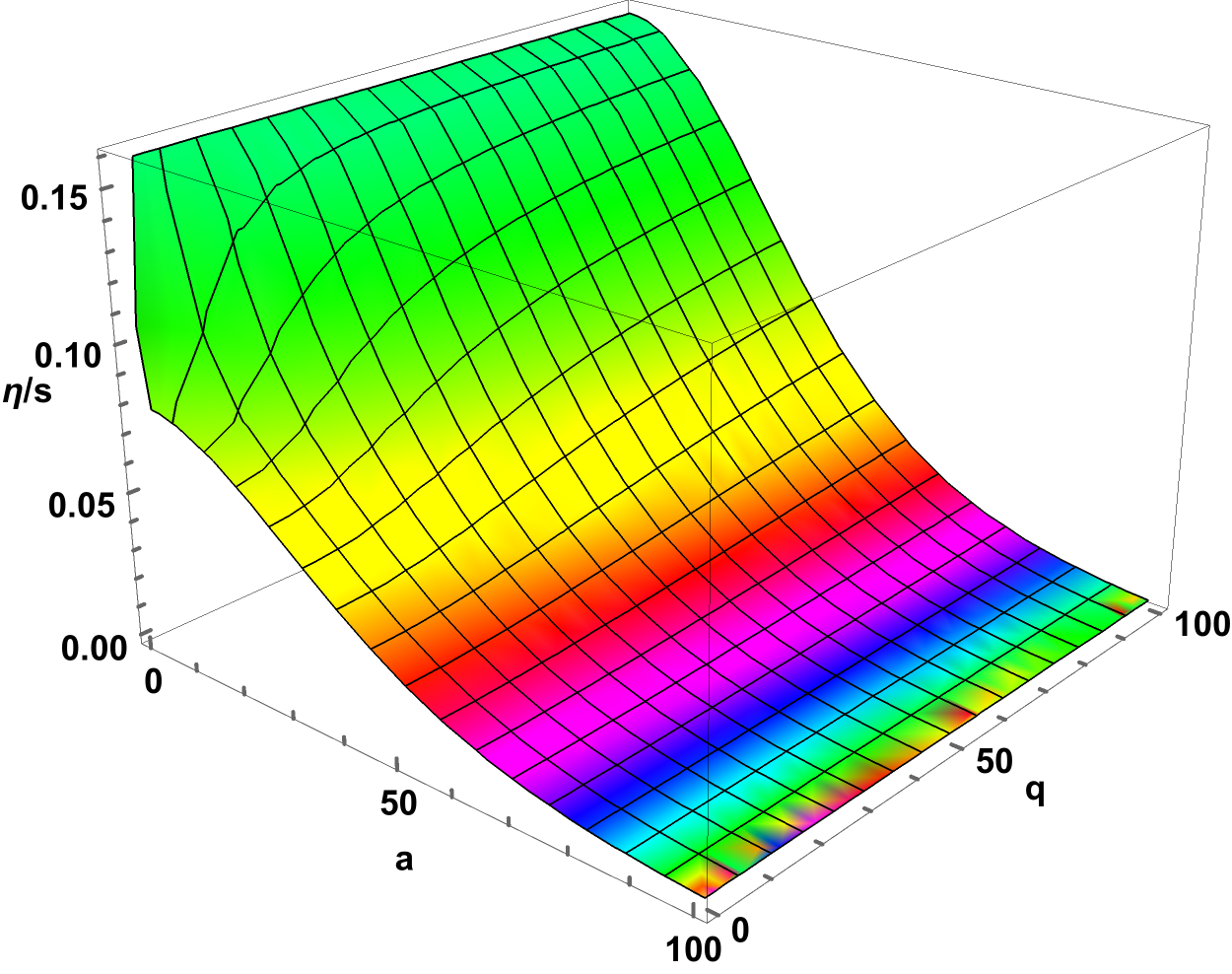}
      \caption{Extremal Kerr-Newman AdS$_4$ black hole $\eta / s$ vs $(a, q)$ for $L = 100$}
      \label{fig:KN BH eta over s VS a q}
      \end{center}
    \end{figure}

Similar to the AdS$_4$ Kerr black hole case, we now consider the wave equation for a minimally coupled scalar field $(\nabla_\mu + i e A_\mu) (\nabla^\mu + i e A^\mu) \Psi = 0$. Following the same strategy, the wave equation is solved in the asymptotic region  and in the near-horizon region separately. Further gluing of the solution in the overlapping region leads to the definition of Green's function in the dual field theory side. As before, we compute the retarded Green's functions for the boundary CFT$_3$ as well as for the near-horizon CFT$_2$. The solutions to the radial part of the wave equation remain the same as the Kerr AdS$_4$ black hole case to the leading order. Hence, the ratio $\eta / s$ formally remains the same for the Kerr-Newman AdS$_4$ black hole
\be
  \frac{\eta}{s} = \frac{r_+^2 L^4 (L^2 - a^2)}{4 \pi (a^2 + r_+^2) (a^2 + L^2)^3}\, ,
\ee
although $r_+$ depends  implicitly also on the charge $q$. One can verify that $\eta / s$  mildly depends on the black hole charge $q$. More importantly, it decreases from the value $1 / (4 \pi)$ as the Kerr-Newman AdS$_4$ black hole angular momentum ($\sim a$) increases, and it vanishes as $a \to L$ in the extremal limit (Fig.~\ref{fig:KN BH eta over s VS a q}).

\par
{\bf Discussion.---}  In this manuscript we have explored the retarded Green's function using the AdS/CFT correspondence in the background of a rotating asymptotically AdS$_4$ black hole.   We have defined two important limits: (i)  one where we consider the full asymptotic region and thus obtained a result compatible with AdS$_4$/CFT$_2$ and (ii) one where we zoom into the near-horizon region and treat that region as the full geometry  whereby we obtained a result compatible with a providing a basis for the Kerr/CFT$_2$ correspondence. We have shown that both limits, (i) and (ii),  are controlled  by one master equation allowing to follow the RG flow in this geometry and establishing  that at both ends one finds backgrounds describing conformal field theories. It is worth remarking that, although the computations presented in this manuscript  focused on $d=3$,  a similar picture should be  valid for $d=4,5,6$ as implied by the unifying picture presented for  rotating,  electrically charged supersymmetric black holes in AdS$_{d+1}$ \cite{David:2020ems}.  In a concrete sense our setup provides a natural UV completion to aspects of the Kerr/CFT$_2$ correspondence by embedding it in AdS$_{d+1}$/CFT$_d$ for $d=3,4,5,6$.  This embedding and its consequent IR decoupling  provide a justification for the Kerr/CFT$_2$ correspondence as a stand-alone conjecture. We explored the shear viscosity to entropy density ratios for the CFT$_3$'s dual to the Kerr-AdS$_4$ and the Kerr-Newman-AdS$_4$ black holes, and found that they vanish for $a \to L$ in the extremal limit. We also established that $\eta/s \sim T^2$ when turning on a small temperature $T$, which precisely matches the result from the entropy production analysis in \cite{Hartnoll:2016tri}.

There are a number of open questions that our work stimulates. It would be natural to discuss the embeddings in other dimensions of the  AdS$_{d+1}$/CFT$_d$ correspondence in more details; we expect, however, that many of the qualitative features of the present work will remain in this more generic situation. Another natural direction is to consider retarded Green's functions of more general operators beyond the scalar ones considered here. In particular, fermionic operators might shed light in the context of potential condensed matter applications as anticipated in \cite{Sonner:2009fk}. Indeed,  for spherically symmetric extremal black holes whose near-horizon region is AdS$_2$, the emergent CFT$_1$  has been shown to realize  mechanisms relevant for strange metal physics \cite{Faulkner:2010zz} (see also \cite{Sachdev:2019bjn}). It would be quite interesting to more precisely pursue the implications of the emergent CFT$_2$ along those lines. Since the asymptotic  boundary of the background \eqref{eq:KerrAdS4BHmetric} is topologically $S^1\times S^2$, our analysis  provides the opportunity to study field theories in curved backgrounds.  Indeed, certain field theoretic behavior relevant for condensed matter applications such as vortex accumulation, can only be studied in field theories on curved backgrounds  \cite{Wiegmann:2013hca}. Similarly, curved backgrounds allow to study the appearance and applications of the gravitational anomaly in quantum Hall effects \cite{Can:2014ota}.  We hope to explore similar applications within our approach in the future.

{\bf Acknowledgments.---}
We would like to thank Daniel Are\'an, Matteo Baggioli, Marina David and Alfredo Gonz\'alez Lezcano for comments. This  work was supported in part by the U.S. Department of Energy under grant DE-SC0007859. J.N. would like to thank Simons Center for Geometry and Physics (SCGP) for warm hospitality during the final stage of this project.


\bibliographystyle{utphys}
\bibliography{Viscosity}

\end{document}